\documentclass{article}
\usepackage[utf8]{inputenc}
\usepackage{authblk}
\usepackage{setspace}
\usepackage[margin=1.25in]{geometry}
\usepackage{graphicx}
\graphicspath{ {./figures/} }
\usepackage{subcaption}
\usepackage{amsmath}
\usepackage{lineno}
\usepackage{textcomp}
\usepackage{multirow}
\usepackage[dvipsnames]{xcolor}
\usepackage{cite}

\title{
Feature-based prediction of properties of cross-linked epoxy polymers by molecular dynamics and machine learning techniques
}

\author[1*]{Sindu B.S.}
\author[2]{Jan Hamaekers}

\affil[1]{Special and Multifunctional Structures Laboratory, CSIR-Structural Engineering Research Centre, Taramani, Chennai, Tamil Nadu, India - 600113.}

\affil[2]{Fraunhofer Institute for Algorithms and Scientific Computing, Schloss Birlinghoven, D-53757 Sankt Augustin, Germany.}

\affil[*]{Corresponding author. Email: sindu@serc.res.in}

\date{\today}

\begin{document}
\maketitle
\begin{abstract}
Epoxy polymers are used in wide range of applications. The properties and performance of epoxy polymers depend upon various factors like the type of constituents and their proportions used and other process parameters. 
The conventional way of designing epoxy polymers may not be fully efficient and labor-intensive and this has lead towards epoxy polymers with limited the performance space due to use of preset mix combinations, compositions and design parameters. 
The conventional way of developing epoxy polymers is usually labor-intensive and may not be fully efficient, which has resulted in 
a limited performance range due to the use of predetermined blend combinations, compositions and development parameters.
Hence, in order to experiment with more design parameters, robust and easy computational techniques need to be established. 
In this regard, an attempt has been  made in this study to develop machine learning (ML) based technique using which the mechanical properties of epoxy polymers can be predicted from their basic structural features.
To this end, we developed and analyzed in this study a new machine learning (ML) based approach to predict the mechanical properties of epoxy polymers based on their basic structural features.
The results from molecular dynamics (MD) simulations have been used to derive the ML model. 
The salient feature of our work is that for the development of epoxy polymers based on EPON-862, several new hardeners were explored in addition to the conventionally used ones.
The salient feature of this work lies in the fact several new curing agents were explored for developing EPON-862 based epoxy polymers in addition to the conventionally used ones.
The influence of additional parameters like the proportion of curing agent used and the extent of curing on the mechanical properties of epoxy polymers were also investigated. This method can further be extended where the epoxy polymer can be designed with desired properties by knowing the structural features of its constituents. 
This method can be further extended by providing the epoxy polymer with the desired properties through knowledge of the structural characteristics of its constituents.
The findings of our study can thus lead toward development of efficient design methodologies for epoxy polymeric systems.          
\end{abstract}%
\providecommand{\keywords}[1]
{
  \small	
  \textbf{\textit{Keywords---}} #1
}
\section{Introduction}
Thermoset epoxy polymers are widely used as matrix/adhesive material in structural, aeronautical and marine applications. They are formed due to reaction between two base monomers (resin and hardener) and the reaction leads to formation of a complex three-dimensional network. Thus, the three-dimensional network and the properties of the resulting polymer depend upon several factors like the type of base monomers used, their composition, parameters under which curing is carried out, extent of curing and the like. Over several decades, the design of epoxy polymers has been carried out using conventional experiments with preset resin/hardener combination and curing process which has resulted in the epoxy polymer space with limited performance characteristics. Hence, in order to expand the epoxy polymer space with high and multi-functional performance characteristics for futuristic applications, advanced techniques should be used to understand the nature/interactions of epoxy polymers at fundamental length scale and identify their correlation with properties/performance at the application length scale.

For this purpose, molecular dynamics (MD) approaches were used, in which the reactions (cross-linking processes) were virtually simulated with different resin/hardener combinations at different curing levels, and their thermo-mechanical properties such as elastic modulus, density, and glass transition temperature (T$_g$) were determined~\cite{orselly2022molecular, odegard2021molecular}.
MD simulations with the {\em ReaxFF}  potential \cite {senftle2016reaxff} were used to simulate the bond dissociation and reorganization phenomena, thus enabling understanding of plastic deformation and fracture process as well as predicting the gel-point, volumetric shrinkage and yield strength \cite{konrad2021molecular, patil2021reactive}. MD simulations were also used to determine the mechanisms of moisture diffusion, thermal conductivity and hygro-thermal degradation in cross-linked epoxy polymers by investigating the inter/intra-molecular interactions (radial distribution function, polar interactions, hydrogen bonding activity, dipole moment fluctuations) \cite{liu2020molecular, sindu2022atomistic, guha2020atomistic}. Moreover, MD simulations were carried out to gain insights on mechanisms of water/ion migration at the composite interfaces (epoxy-cement, epoxy-fiber) as in the case of chemical attack and to evaluate their influence on the inter-facial integrity \cite{tam2019molecular, hou2020nanoscale}. Furthermore, it was demonstrated that MD can also be used for design, property prediction and understanding interactions between various constituents in nano-engineered/smart/multi-functional epoxy polymer systems \cite{jian2021molecular, yang2019molecular, sindu2015evaluation}.
While MD helps in simulating polymer systems with new combinations and new process parameters, and also contributes to the understanding of the fundamental phenomena that occur in various macroscopic processes, the method is limited because it is time consuming and computationally intensive.

In recent times, machine learning (ML) techniques were also applied for design, optimization and property prediction of epoxy polymers. For example, ML based active learning and Bayesian optimization were used to predict and optimize the process parameters for development of epoxy polymers with high adhesive strength. In particular, this approach lead towards significantly less experimental trials (32 as against 256 possible combinations) to design epoxy polymer with very high adhesive strength \cite{pruksawan2019prediction}. A similar approach was also used for the design of multi-component, bio-based epoxy polymers with high and low T$_g$ \cite{albuquerque2023designing}.
In addition, artificial neural networks (ANN) were used to develop shape memory epoxy polymers (SMPs) with optimized performance characteristics (T$_g$, flexural strength, strain fixation rate, and strain recovery rate), where the input of the ML model are the constituents of the epoxy polymer and the output is the polymer property \cite{liu2022performance}.
A method involving Raman spectroscopy in conjunction with ML techniques (random forest and partial least square regression) was used to determine the optimum molar composition of epoxy polymers which is one of the most determining factor that dictates the performance and properties of epoxy polymers \cite{guan2019rapid}.  ML based technique was also applied to predict and quantify the effects of environmental factors (thermal exposure - different temperatures and time periods) on the degradation in properties of epoxy polymers~\cite{doblies2019prediction}. Recently, BigSMILES notation (string based representation of molecules which can also support stochastic molecular structures) was employed as an input for convolutional neural network (CNN) models for polymer property prediction. A CNN model was developed which can predict both the T$_g$ and recovery stress of SMP. The model was further used to design new SMPs which had higher recovery stress than the known ones \cite{yan2021machine}. ML also proved to be a powerful tool in understanding correlations between fundamental structural features/genes and properties of epoxy polymers. ML ensemble model was used to predict the T$_g$ of epoxy polymers with different resin/hardener combinations (16 resins and 19 hardeners) using the molecular descriptors determined from Mordred \cite{moriwaki2018mordred} and RDKit \cite{RDKit} in combination with principal component analysis (PCA). Using this approach, it has been identified that T$_g$ is expected to increase for smaller, less polarizable and more hydrophobic resins \cite{meier2022modeling}. A ML assisted materials genome approach combining concepts of graph convolutional networks (attention- and gate- augmented), classical gel theory and transfer learning was used for design of epoxy polymers with high mechanical properties (elastic modulus, tensile strength and toughness) based on experimental results collected for epoxy polymers and polyamides from which the gene structures affecting these properties were also identified \cite{hu2022machine}.
While ML is promising for designing new polymers with unusual properties and understanding relationships that are not possible with conventional techniques, to create an accurate model, it needs to be trained with more data, which in the case of epoxy polymers is limited.

In order to overcome the limitations of individual techniques and to utilize their benefits, in recent times, appropriate combinations of these two methods are being used, wherein, MD is used to simulate new polymeric systems with different blends and process parameters, the results from MD is used for developing ML models to get deeper insights. For example, MD was used to simulate epoxy polymers (30 different types) with different resin/hardener/toughening agent combinations including different proportions and their results were further used to develop ML model for optimizing the constituents to design high performance epoxy polymer with high elastic modulus, ultimate tensile strength, density and T$_g$ \cite{jin2020composition}. Similarly, MD was used in conjunction with ML for predicting the properties of self-healing epoxy polymers and for optimizing their constituents \cite{luo2021properties}. These methods were further extended to determine the correlation between constituent and polymer property using Pearson coefficient to identify the constituent responsible for each of the property (cohesive energy density, elastic modulus and T$_g$) \cite{choi2022predicting}. Coarse grained (CG) simulations and ML based methods were used to predict the viscosity of different epoxy polymer clusters (categorized using K-means clustering) with different diluent proportions and at various temperatures \cite{qiu2022highly}. While the combination of both methods seem more promising, very limited studies are carried out using these techniques together and are mostly limited to composition optimization with a preset resin/hardener combination or identifying the constituent-property relationship within the limited blend combinations inspite of wide range of possibilities.

With this in mind, this study will use MD simulations to simulate a wide range of polymers by extending the conventional blend combinations and employ ML techniques to gain deeper insights into the parameters affecting the mechanical properties of these epoxy polymers.
We generate several epoxy polymer systems with different types of hardeners (aromatic rings/aliphatic rings/aliphatic chains), varying proportions of hardeners and also from partially cured to maximum cured systems, cf.\ Section 2.1-2.2. Then, subject these generated systems to strain simulations, cf.\ Section 2.3 and from their stress-strain response, key mechanical properties like yield strength and elastic modulus are determined, cf.\ Section 3. Further, ML based techniques are employed to predict the properties of different epoxy polymer systems wherein the structural features of the hardeners (since same resin is used for all cases, the features of resin was not considered) are used as input, cf.\ Section 4. Finally, the structural feature-property correlation is established for thermoset epoxy polymers. The findings of this study would enable efficient design of epoxy polymers with desired mechanical properties by merely selecting constituents with relevant structural features. This study has also paved the way for expanding the epoxy polymer space with more types of combinations in addition to the conventional ones.

\section{Molecular dynamics simulations} \label{sec:2}
\begin{figure}[hbt!]
    \centering
    \includegraphics[width=0.50\textwidth]{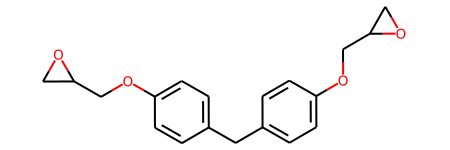}
    \caption{Epoxy resin considered in this study (Diglycidyl Ether of Bisphenol-F (EPON 862)).}
    \label{fig:1}
\end{figure}

In this study, MD simulations are carried out on different epoxy systems formed by atomistic cross-linking procedure. All the epoxy systems have same resin, Diglycidyl Ether of Bisphenol-F (EPON 862) (as shown in Figure \ref{fig:1}) but different hardeners. A total of 11 amine-based hardeners (as shown in Table \ref{tab:1}) is considered with different structural configurations (aromatic amines/aliphatic amines, primary/secondary amines). All simulations are carried out using a commercial software (QuantumATK) \cite{schneider2017atk} with OPLS-AA (Optimized Potentials for Liquid Simulations) potential \cite{jorgensen1996development} built in Tremolo-X Calculator \cite{schneider2017atk}. The functional form of OPLS-AA potential is presented in equation \eqref{eq1}, in which the total potential energy of the system is the summation of contributions due to bonded (bond stretching, angle bending, torsional and inversion terms) and non-bonded (Lennard-Jones and electrostatic) interactions.
\begin{equation}\label{eq1}
\begin{split}
    E(x) = \sum_{i}^{bonds} k_i(r_i-r_0)^2 + \sum_{i}^{angles} k_i(\theta_i-\theta_0)^2\ + \sum_{i}^{torsions} \sum_{n=1}^{3} k_{i,n}(1-cos(n\phi_i-\delta_{i,n})) \\+ \sum_{i}^{inversion} k_i(1-cos(n\chi_i)) + \sum_{ij}^{atoms}4\varepsilon_{ij}\left[\left(\frac{\sigma_{ij}}{r_{ij}}\right)^{12}-\left(\frac{\sigma_{ij}}{r_{ij}}\right)^6 \right] + \sum_{ij}^{atoms}\frac{q_iq_j}{4\pi\epsilon_0r_{ij}}
\end{split}
\tag{1}
\end{equation}
where \textsl{r}, $\theta$, $\phi$ \;and $\chi$ \;denote bond distances, angles, torsions and inversions respectively. \newline

\begin{table}[hbt!]
    \caption{Hardeners considered in this study.}
    \centering
    \begin{tabular} {llc}
            \hline
            Hardener & Molecular conf. & Type\\
            \hline
            Diethyl toluene diamine (DETDA) & \includegraphics[scale=0.15]{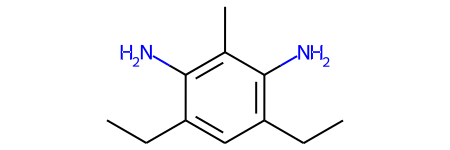} & \multirow {3}{*} {Aromatic ring}\\
            Phenylene diamine (PPD) & \includegraphics[scale=0.15]{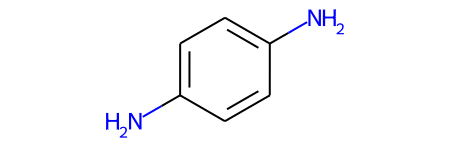} &\\
            Diamino diphenyl methane (DDM) & \includegraphics[scale=0.15]{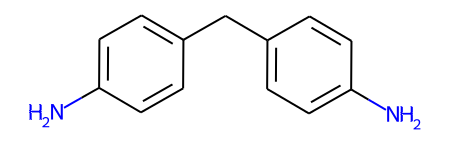} &\\
            \hline
            Diamino cyclohexane (DACH) & \includegraphics[scale=0.15]{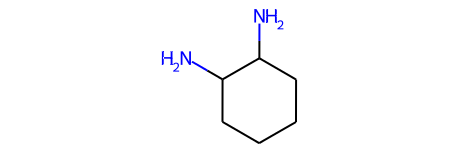} & \multirow {3}{*} {Aliphatic ring}\\
            Aminoethyl piperazine (PIPER) & \includegraphics[scale=0.15]{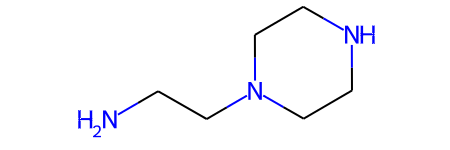} &\\
            Isophorone diamine (IPDA) & \includegraphics[scale=0.15]{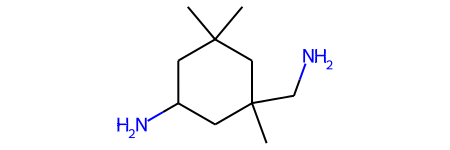} &\\
            \hline
            Triethylene tetramine (TETA) & \includegraphics[scale=0.15]{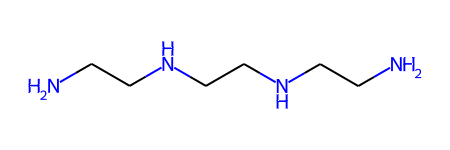} & \multirow {5}{*} {Aliphatic chain}\\
            Diethylene triamine (DETA) & \includegraphics[scale=0.15]{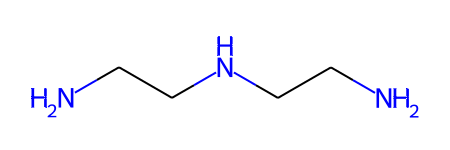} &\\
            Trimethyl hexamethylene diamine (TMD) & \includegraphics[scale=0.15]{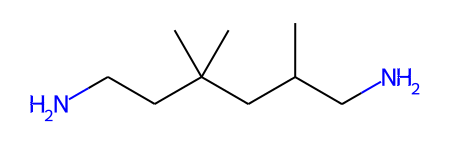} &\\
            Trioxa tridecane diamine(TTD) & \includegraphics[scale=0.15]{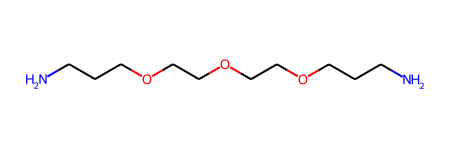} &\\
            Polyoxypropylene diamine (Jeffamine) & \includegraphics[scale=0.15]{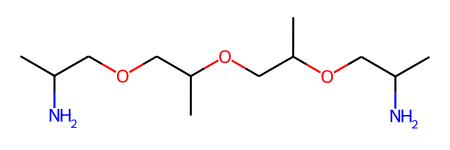} &\\
            \hline
            \end{tabular}
    \label{tab:1}
\end{table}

\subsection{Construction of simulation box} \label{subsec:2.1}
Firstly, the molecular configuration of monomers (resin/hardener) is generated using SMILES (simplified molecular-input line-entry system) \cite{weininger1988smiles}. The monomers of resin and hardener are then randomly inserted into the simulation box of size 65 x 65 x 65 Å (number of atoms $\approx$ 15000 atoms) with a packing density of 0.6 g/cc using Packmol tool \cite{martinez2009packmol}. The number of resin and hardener molecules to be inserted into the simulation box is calculated using the stoichiometric ratio, molecular mass of the monomers and the packing density. The tolerance (intermolecular distance) and buffer (to the cell vectors) during packing is considered as 2 Å. The packed simulation box  is then subjected to geometry optimization using LBFGS algorithm with a force tolerance of 0.05 eV/Å. The optimization is carried out in 500 steps with a step size of 0.2 Å. The optimized simulation box is further equilibrated in three stages (NVT @ 300 K, NPT @ 1 bar, 300 K and NPT @ 1 bar, 480 K). All simulations are carried out with a timestep of 1 fs. Berendsen thermostat and barostat are used in the first two stages of equilibration with 50000 steps while Martyna-Tobias-Klein barostat is used in the last stage with 100000 steps where high temperature NPT simulations are taking place.
\subsection{Cross-linking procedure} \label{subsec:2.2}
\begin{figure}[hbt!]
    \centering
    \begin{subfigure}{0.45\textwidth}
        \includegraphics[height=0.2\textheight]{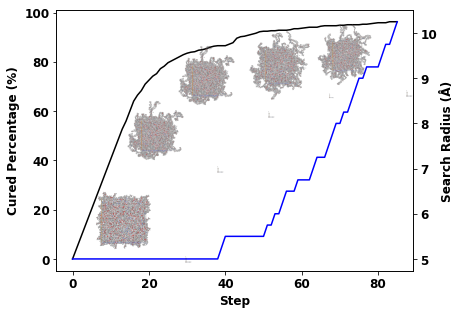}
        \caption{\label{fig:2a}}
    \end{subfigure}
    \begin{subfigure}{0.45\textwidth}
        \includegraphics[height=0.2\textheight]{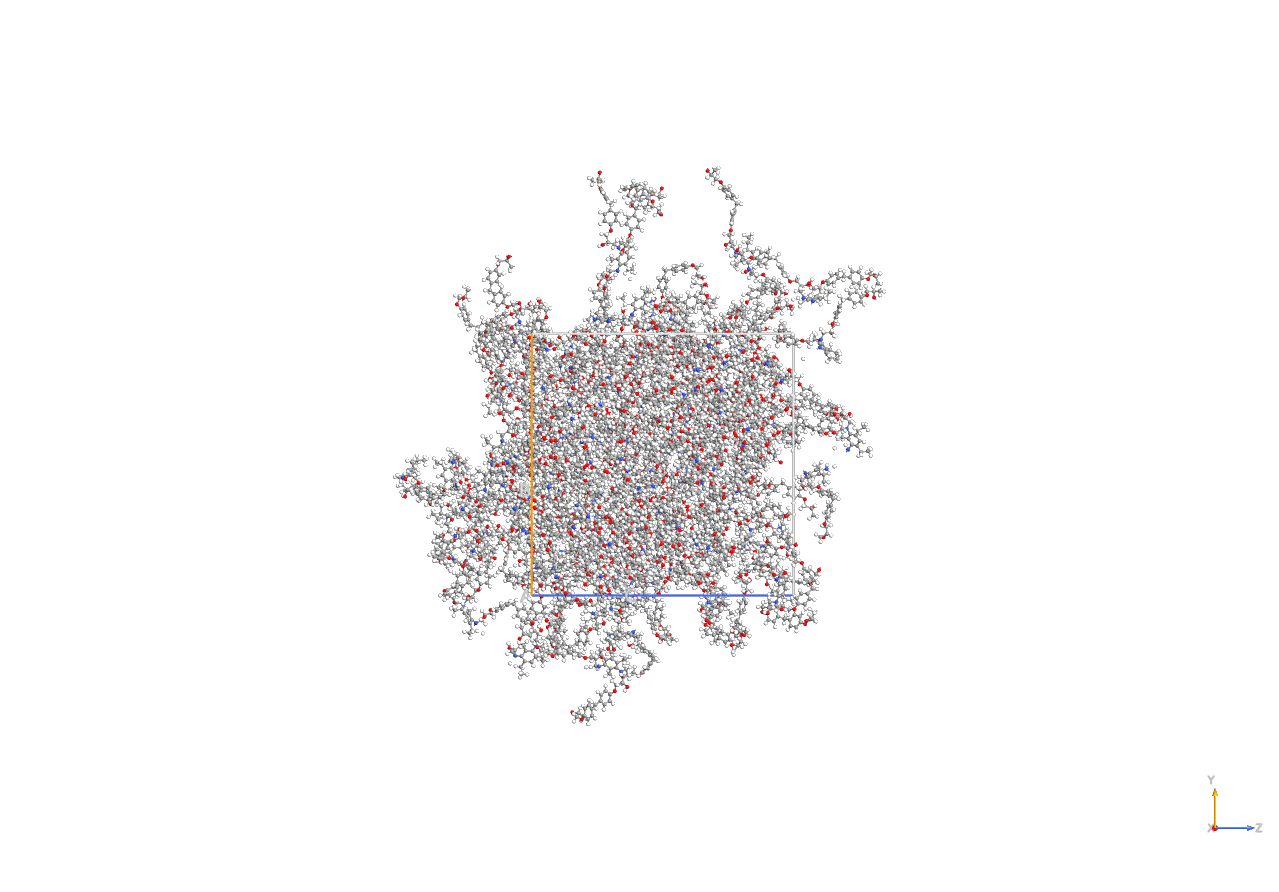}
        \caption{\label{fig:2b}}
    \end{subfigure}
    \caption{(\subref{fig:2a}) Evolution of the cross-linking process (inscribed figures shows the simulation box at different degrees of curing). (\subref{fig:2b}) Fully cross-linked structure.}
    \label{fig:2}
\end{figure}

The equilibrated structure is then subjected to a cross-linking process. During this process, reactions are simulated between the epoxide and amine (aromatic/aliphatic) groups in the resin and hardener respectively. More details about the cross-linking process can be found elsewhere \cite{sindu2022atomistic, smidstrup2019quantumatk, varshney2008molecular, kallivokas2019molecular}.
Bonds are formed between these reaction groups when they are at a certain distance.
The bond search radius is gradually increased from 5 Å to 10 Å with a step size of 0.25 Å. During each cycle, the system is equilibrated using NVT (10000 steps) followed by NPT simulations (10000 steps). The evolution of the curing process along with the cross-linked simulation systems is presented in Figure \ref{fig:2}. Since high temperature leads to better cross-linking degree, the whole process is simulated at 480 K. The simulation box with the cross-linked structure is further equilibrated to bring the polymeric system to the correct density (1.13 g/cc). As the relaxation of polymeric system usually takes place at a longer timescales, the process is accelerated using 21-step equilibration process (as suggested by \cite{abbott2013polymatic}) using cycles of high temperature and high pressure simulations and annealing until the desired temperature and pressure is reached.

\subsection{Stress-strain simulations}\label{subsec:2.3}
\begin{figure}[hbt!]
    \centering
    \includegraphics[width=0.50\textwidth]{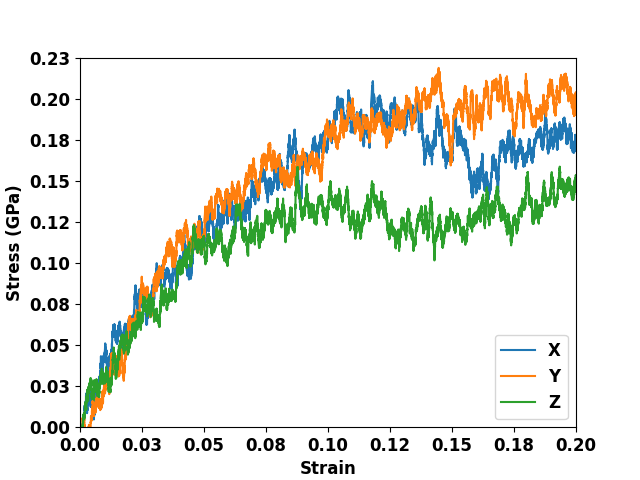}
    \caption{Stress-strain response of DGEBF-DETDA obtained from strain simulations in three orthogonal directions (X,Y,Z).}
    \label{fig:3}
\end{figure}

The final equilibrated structure is then used for the production stage where the actual simulations are performed to obtain the material properties. Strain is applied to the simulation box at a constant rate and the corresponding stress is measured and from the stress-strain relationship, properties like elastic modulus and yield strength are determined \cite{griebel2004molecular, sindu2020molecular}. Since epoxy polymers are isotropic materials, the strain simulation is carried out in all three directions (as shown in Figure \ref{fig:3} and the average value is taken as the computed material property. The elastic modulus is calculated from the slope of the stress-strain curve when the material is still in elastic stage (a strain level of 0.025). The yield strength is taken as the stress at which the slope of the stress-strain curve changes to negative. In order to get the properties, five different samples (structures) are generated for each case and the results are averaged (from all fives cases and in each case three different directions).

\section{Mechanical properties of epoxy polymers from MD simulations} \label{sec:3}
\begin{figure}[hbt!]
    \centering
    \begin{subfigure}{0.45\textwidth}
        \includegraphics [width=0.95\textwidth]{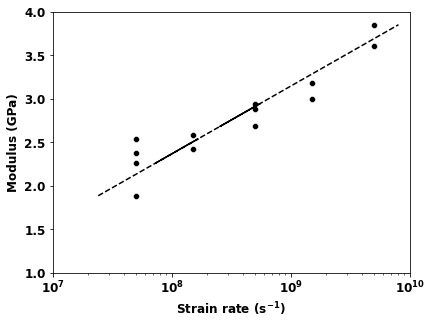}
        \caption{\label{fig:4a}}
    \end{subfigure}
    \begin{subfigure}{0.45\textwidth}
        \includegraphics [width=0.95\textwidth]{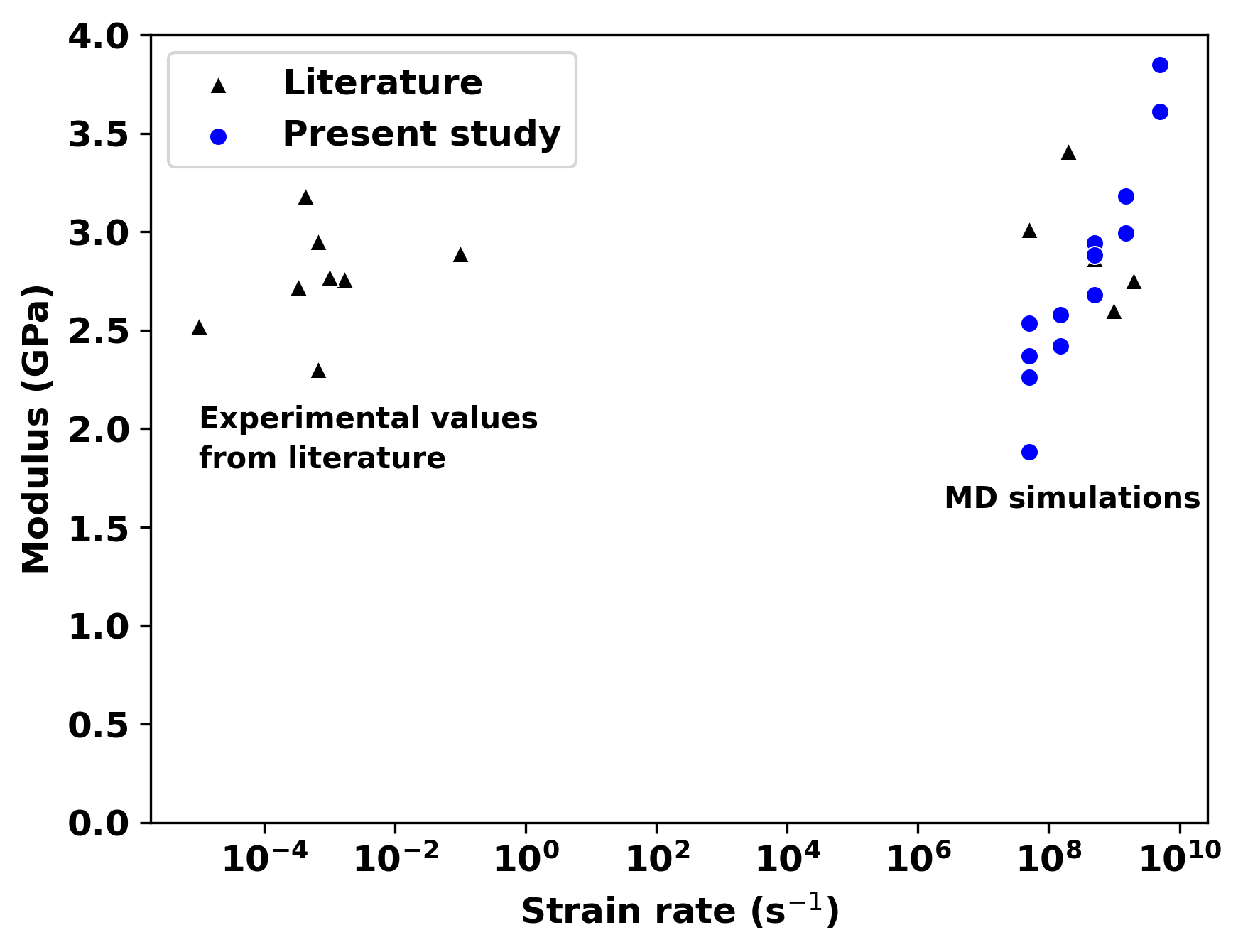}
        \caption{\label{fig:4b}}
    \end{subfigure}
    \begin{subfigure}{0.45\textwidth}
        \includegraphics [width=0.95\textwidth]{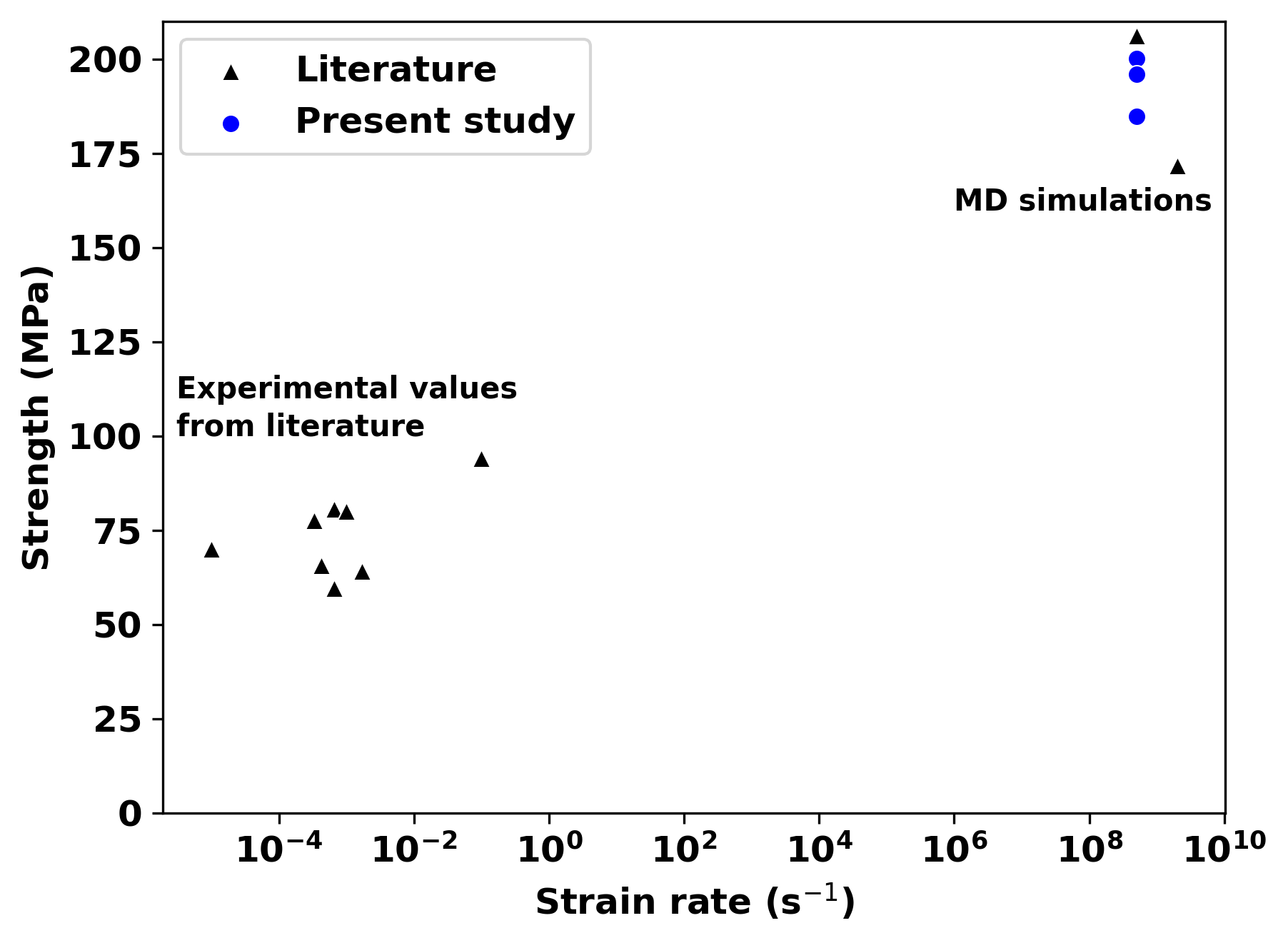}
        \caption{\label{fig:4c}}
    \end{subfigure}
    \caption{(\subref{fig:4a}) Influence of strain rate on MD results. MD results compared with experimental results from literature \cite{patil2021reactive, kallivokas2019molecular,  sun2008mechanical, bajpai2019tensile, shinde2014comparison, littell2008measurement, king2015mechanical, gollins2014comparison, li2011molecular, kelkar2011prediction, wan2021thermal, li2015evolution}: (\subref{fig:4b}) Elastic modulus. (\subref{fig:4c}) Yield strength.}
    \label{fig:4}
\end{figure}

The mechanical properties of epoxy polymers determined from MD simulations are compared with those of the experimental results reported in the literature \cite{patil2021reactive, kallivokas2019molecular,  sun2008mechanical, bajpai2019tensile, shinde2014comparison, littell2008measurement, king2015mechanical, gollins2014comparison, li2011molecular, kelkar2011prediction, wan2021thermal, li2015evolution, fard2012characterization, jakubinek2018nanoreinforced}. Since strain rate plays a crucial role on the obtained results, simulations (of a particular combination, DGEBF-DETDA) are carried out with five different strain rates (0.005~ps$^{-1}$, 0.0015~ps$^{-1}$, 0.0005~ps$^{-1}$, 0.00015~ps$^{-1}$, 0.00005~ps$^{-1}$). The elastic modulus of DGEBF-DETDA system from different strain rate simulations is presented in Figure \ref{fig:4a} where it can be found that a linear relationship exists between the elastic modulus and strain rate. In general, the strain rates used in MD simulations are very high in comparison to that used in actual experiments (0.001~s$^{-1}$).
The comparisons of elastic modulus and yield strength obtained from MD simulations and those reported in experiments are presented in Figures \ref{fig:4b} and \ref{fig:4c} respectively and in Table \ref{tab:2}.
It can be observed that in case of elastic modulus, except for the case with the strain rate of 0.005 ps$^{-1}$, all values from simulation fall within the  values (experimental and computational) reported in literature. Simulations with lower strain rates (0.00015~ps$^{-1}$, 0.00005~ps$^{-1}$) typically require longer simulation time until yielding takes place (to determine yield strength). Hence, in order to strike a balance between accuracy and computational demand, a strain rate of 0.0005~ps$^{-1}$ is used for further simulations. The yield strength obtained from the present study also falls well within the range of values reported (from MD simulations) in literature. However, the yield strength obtained from MD simulations is found to be higher (3 times) than the values reported in literature from experimental investigations. This phenomenon is also found to be logical in the sense that, in usual, the strength is strongly dependent on strain rate, and higher the strain rate, the strength is over-predicted. Further, the influence of additional parameters like the degree of curing and the type of hardener used on the mechanical properties like elastic modulus and yield strength is determined.

\begin{table}[hbt!]
    \caption{Validation of mechanical properties, i.e.\ elastic modulus (EM) and tensile yield strength (TS), with experimental results.}
    \centering
    \begin{tabular} {llllll}
            \hline
            Epoxy polymer & Strain rate & EM (GPa) & TS (MPa) & Method & Ref.\\
            \hline
            DGEBF-DETDA & 5.08 mm/min. & 2.76$\pm$0.01 & 64.1$\pm$5.6 & Expt. & \cite{sun2008mechanical}\\
            DGEBF-DETDA & 2 mm/min. & 2.95$\pm$0.02 & 80.6$\pm$3.1 & Expt. & \cite{bajpai2019tensile}\\
            DGEBF-DETDA & 1.27 mm/min. & 3.18$\pm$0.12 & 65.63$\pm$7.37 & Expt. & \cite{shinde2014comparison}\\
            DGEBF-DETDA & 1x10$^{-5}$s$^{-1}$ & 2.52 & 70 & Expt. & \cite{littell2008measurement}\\
            DGEBF-DETDA & 1x10$^{-3}$s$^{-1}$ & 2.77 & 80 & Expt. & \cite{littell2008measurement}\\
            DGEBF-DETDA & 1x10$^{-1}$s$^{-1}$ & 2.89 & 94 & Expt. & \cite{littell2008measurement}\\
            DGEBF-DETDA & 1 mm/min. & 2.72$\pm$0.04 & 77.6$\pm$0.9 & Expt. & \cite{king2015mechanical}\\
            DGEBF-DETDA & 2 mm/min. & 2.3 & 59.7 & Expt. & \cite{gollins2014comparison}\\
            DGEBF-DETDA & 5x10$^{8}$s$^{-1}$ & 2.86 & 206.16 & MD & \cite{li2011molecular}\\
            DGEBF-DETDA & 5x10$^{5}$s$^{-1}$ & 4.60 & - & MD & \cite{kelkar2011prediction}\\
            DGEBF-DETDA & 5x10$^{7}$s$^{-1}$ & 3.01 & - & MD & \cite{wan2021thermal}\\
            DGEBF-DETDA & 2x10$^{9}$s$^{-1}$  & 2.75 & 171.72 & MD & \cite{li2015evolution}\\
            DGEBF-DETDA & 1x10$^{9}$s$^{-1}$ & 2.60 & - & MD & \cite{kallivokas2019molecular}\\
            DGEBF-DETDA & 2x10$^{8}$s$^{-1}$  & 3.41 & - & MD & \cite{patil2021reactive}\\
            DGEBF-DETDA & 5x10$^{8}$ s$^{-1}$ & 2.83$\pm$0.14 & 604.54$\pm$12.24 & MD & Present work\\
            \hline
            DGEBF-DETA & 5.9x10$^{-5}$ s$^{-1}$ & 2.97 & 72 & Expt. & \cite{fard2012characterization}\\
            DGEBF-DETA & 4.93X10$^{-4}$ s$^{-1}$ & 3.08 & 81 & Expt. & \cite{fard2012characterization}\\
            DGEBF-DETA & 0.5 mm/min. & 2.29$\pm$0.07 & 62$\pm$6 & Expt. & \cite{jakubinek2018nanoreinforced}\\
            DGEBF-DETA & 5x10$^{8}$ s$^{-1}$ & 2.69$\pm$0.18 & 621.26$\pm$12.45 & MD & Present work\\
            \hline
    \end{tabular}
    \label{tab:2}
\end{table}

\subsection{Influence of stoichiometric ratio and curing percentage} \label{subsec:3.1}
\begin{figure}[hbt!]
    \centering
    \begin{subfigure}{0.45\textwidth}
        \includegraphics[width=0.95\textwidth]{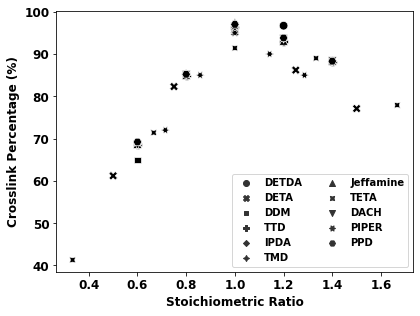}
        \caption {\label{fig:5a}}
    \end{subfigure}
    \begin{subfigure}{0.5\textwidth}
       \includegraphics[width=0.95\textwidth]{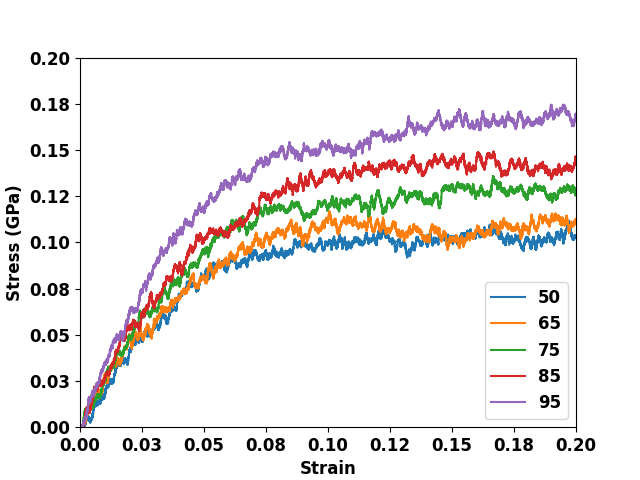}
        \caption {\label{fig:5b}}
    \end{subfigure}
    \caption{(\subref{fig:5a}) Influence of stoichiometric ratio on the final cross-linked percentage of different epoxy polymer systems. (\subref{fig:5b})  Influence of curing percentage on the stress-strain response of epoxy polymers (a typical case of DGEBF-DETDA presented).}
    \label{fig:5}
\end{figure}

\begin{figure}[hbt!]
    \centering
    \begin{subfigure}{0.75\textwidth}
        \includegraphics [width=0.95\textwidth]{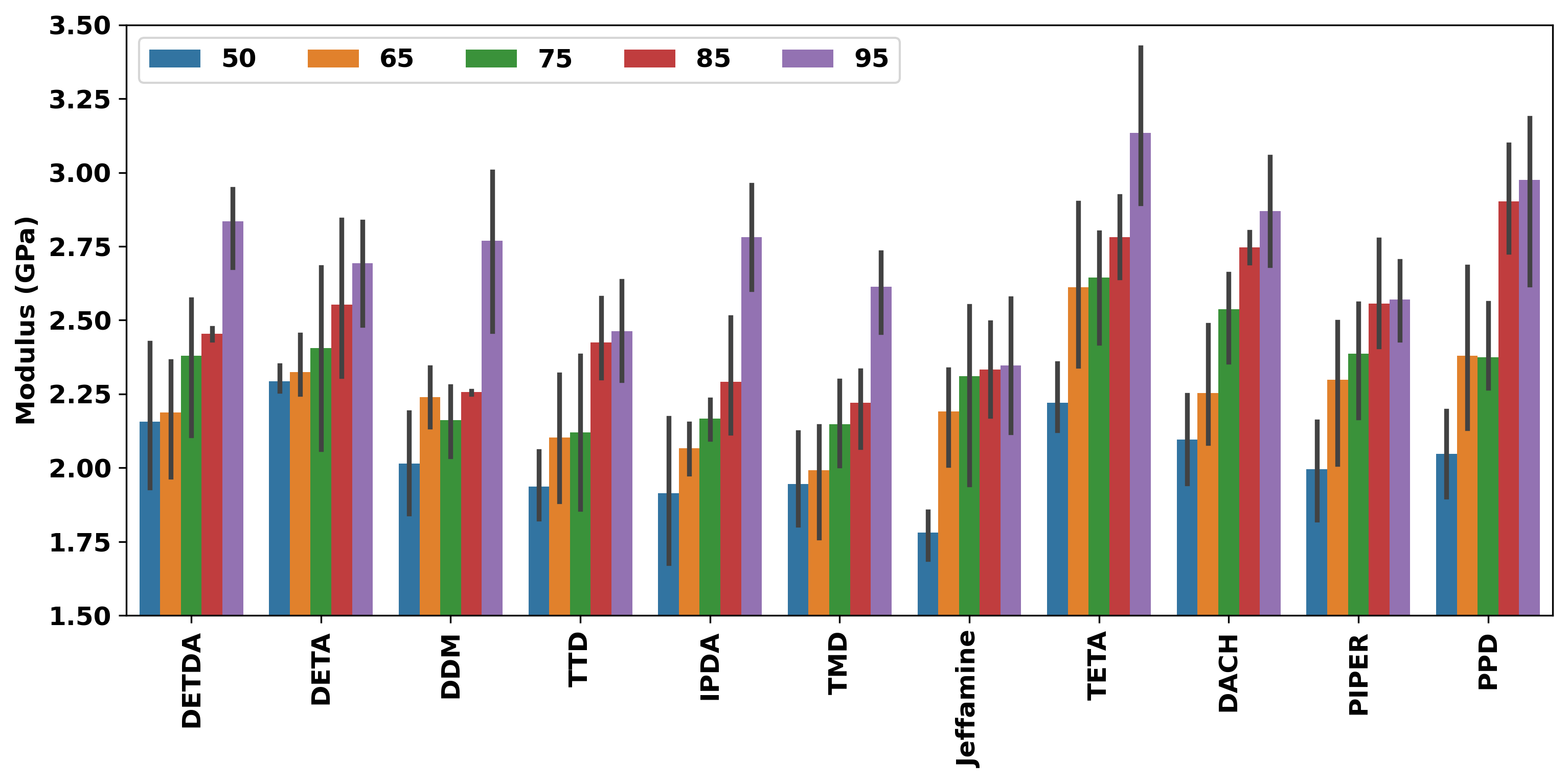}
        \caption{\label{fig:6a}}
    \end{subfigure}
    \begin{subfigure}{0.75\textwidth}
        \includegraphics [width=0.95\textwidth]{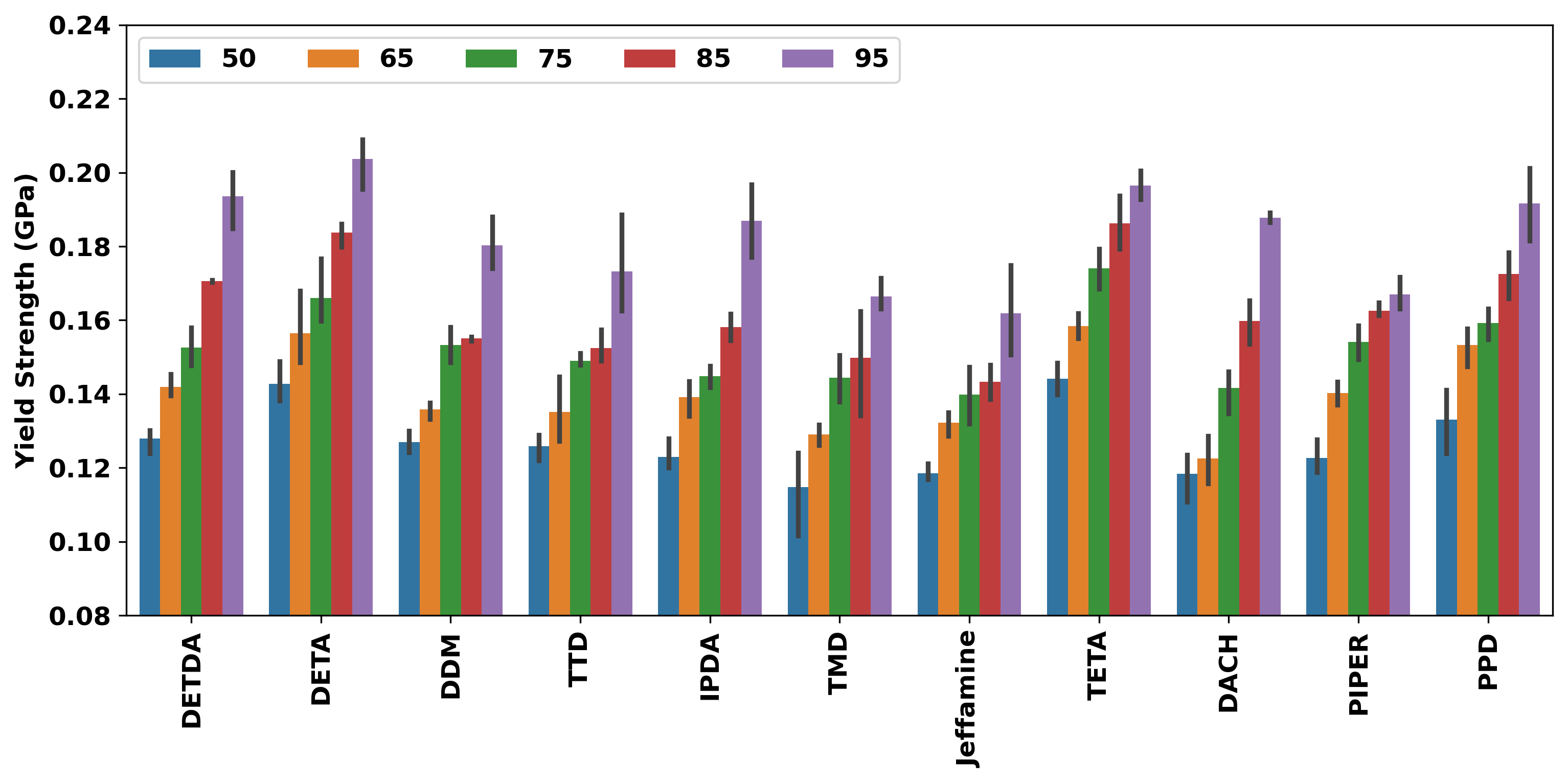}
        \caption{\label{fig:6b}}
    \end{subfigure}
    \caption{Influence of curing percentage on the mechanical properties of epoxy polymers: (\subref{fig:6a}) Elastic modulus. (\subref{fig:6b}) Yield strength.}
    \label{fig:6}
\end{figure}

Since epoxy polymers are formed by reaction between resin and hardener, inclusion of the right quantity of resin and hardener is very important for proper reaction to take place. Each primary amine (-NH$_2$) of the hardener can react with two epoxide groups and each secondary amine (-NH) can react with one epoxide group of the resin. So the ratio of resin and hardener used for preparation of epoxy polymers should be such that the mix contains right proportion of amine/epoxide groups (stoichiometric ratio = 1). If less quantity of resin is present in the mix, the stoichiometric ratio is less than one and if more quantity of resin is present, the stoichiometric ratio is greater than one. The stoichiometric ratio greatly affects the curing percentage (reaction percentage) of the final product, epoxy system. The maximum curing percentage achieved in different epoxy polymer systems with stoichiometric ratio ranging from 0.4 to 1.7 is presented in Figure \ref{fig:5a}. The curing percentage, in turn, affects the mechanical properties of epoxy polymers. The stress-strain response of DGEBF-DETDA at different levels of curing is presented in Figure \ref{fig:5b}. It can be observed that both the elastic modulus and yield strength are affected by the level of curing. The results of all epoxy combinations considered in this study at different levels of curing are presented in Figures \ref{fig:6a} and \ref{fig:6b}, respectively. It can be observed that the elastic modulus of the considered epoxy polymer systems is found to decrease by 15-31\% when the curing percentage is only 50\% as against the fully cured system;  however, the yield strength is found to decrease by 26-36\%.

\subsection{Influence of the type of hardener} \label{subsec:3.2}
\begin{figure}[hbt!]
    \centering
    \begin{subfigure}{0.45\textwidth}
        \includegraphics [width=0.95\textwidth]{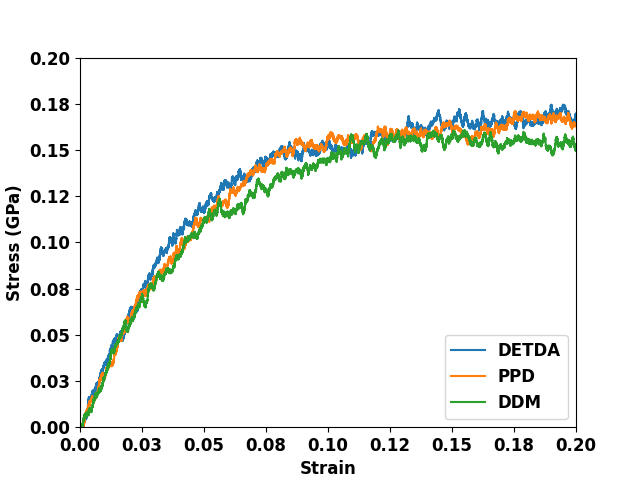}
        \caption{\label{fig:7a}}
    \end{subfigure}
    \begin{subfigure}{0.45\textwidth}
        \includegraphics [width=0.95\textwidth]{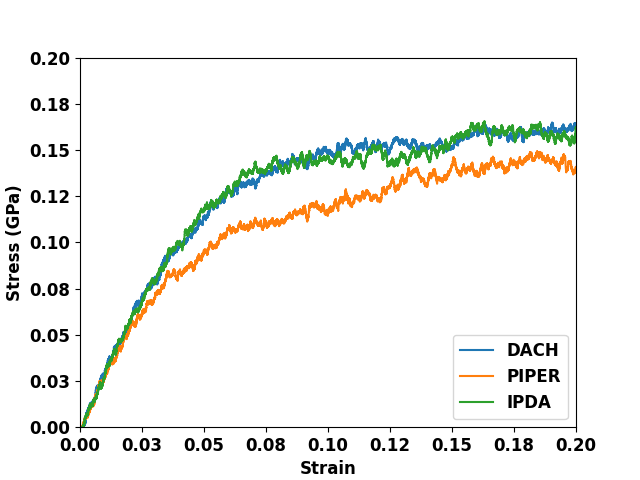}
        \caption{\label{fig:7b}}
    \end{subfigure}
    \begin{subfigure}{0.45\textwidth}
        \includegraphics [width=0.95\textwidth]{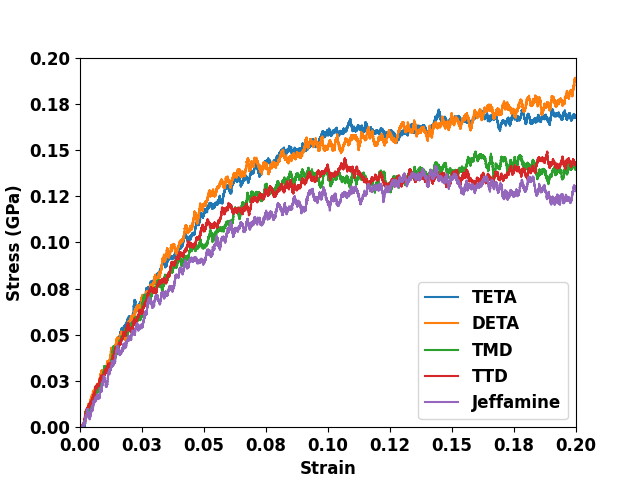}
        \caption{\label{fig:7c}}
    \end{subfigure}
    \caption{Influence of the type of hardener: (\subref{fig:7a}) Hardener with aromatic rings. (\subref{fig:7b}) Hardener with aliphatic rings. (\subref{fig:7c}) Hardener with aliphatic chains.}
    \label{fig:7}
\end{figure}

The stress-strain response of fully cured (\>95\%) epoxy polymers pertaining to different hardener classes, i.e., hardeners with aromatic rings, aliphatic rings and aliphatic chains are presented in Figures \ref{fig:7a}, \ref{fig:7b} and \ref{fig:7c} respectively. In the case of hardeners with aromatic rings (Figure \ref{fig:7a}), DDM is found to possess less mechanical properties when compared to DETDA and PPD. A simple visual correlation of the atomic structure (presented in Table \ref{tab:1}) reveals that DDM possess two aromatic rings whereas DETDA and PPD possess only one aromatic ring. In the case of hardeners with aliphatic rings (Figure \ref{fig:7b}), DACH and IPDA are found to perform better than PIPER. The visual interpretation from their atomic structures reveals that while DACH and IPDA possess two primary amines, PIPER possess only one primary and one secondary amine that are accessible for reaction. Similarly, in the case of hardeners with aliphatic chains (Figure \ref{fig:7c}), TETA and DETA perform better than the other three hardeners, which can also be attributed to the presence of additional secondary amines in those two cases. From the above findings, it can be concluded that there exists a correlation between atomic structure/feature with its mechanical properties. Hence, we further use ML based techniques to predict the properties of epoxy polymers from it's constituent atomistic features.

\section{Machine learning to predict mechanical properties from atomistic features} \label{sec:4}
In order to predict the mechanical properties of epoxy polymers from the atomistic structural features, we employ machine learning techniques. For each of the epoxy combinations, the following features of the hardener are extracted from cheminformatics tool, RDKit (the features of resin are not considered since same resin is used in all cases):
\begin{itemize}
    \item molecular weight,
    \item number of carbon atoms/oxygen atoms/SP3 hybridized carbon atoms,
    \item number of rotatable bonds,
    \item number of aromatic rings/aliphatic rings,
    \item number of secondary amines/primary amines,
    \item number of radical electrons/valence electrons.
\end{itemize}
We apply a gaussian process regression (GPR) approach to develop a predictive model for the mechanical properties (elastic modulus and yield strength) based on these features, where we label the data using the results of appropriate MD simulations. Note here that GPR is a non-parametric, kernel-based, and probabilistic method that is effective for small datasets with a limited number of features \cite{rasmussen2006gaussian}.
In total we use 55 cases, which consist of epoxy polymers with different hardeners and curing degrees. 
To ensure compatibility between the different feature ranges, we normalize the features to have zero mean and unit variance.

Initially, we use a feature selection technique called sequential backward selection (SBS) to identify the most important structural features for predicting the elastic modulus and yield strength. In each trial, we sequentially remove the features with the least relevance from the initial feature list. This reduces the computational demand without sacrificing prediction accuracy. Then, we evaluate the accuracy of the GPR model using the coefficient of determination ($R^2$ score) as the scoring parameter for SBS.

\begin{figure}[hbt!]
    \centering
    \begin{subfigure}{0.45\textwidth}
        \includegraphics [width=0.95\textwidth]{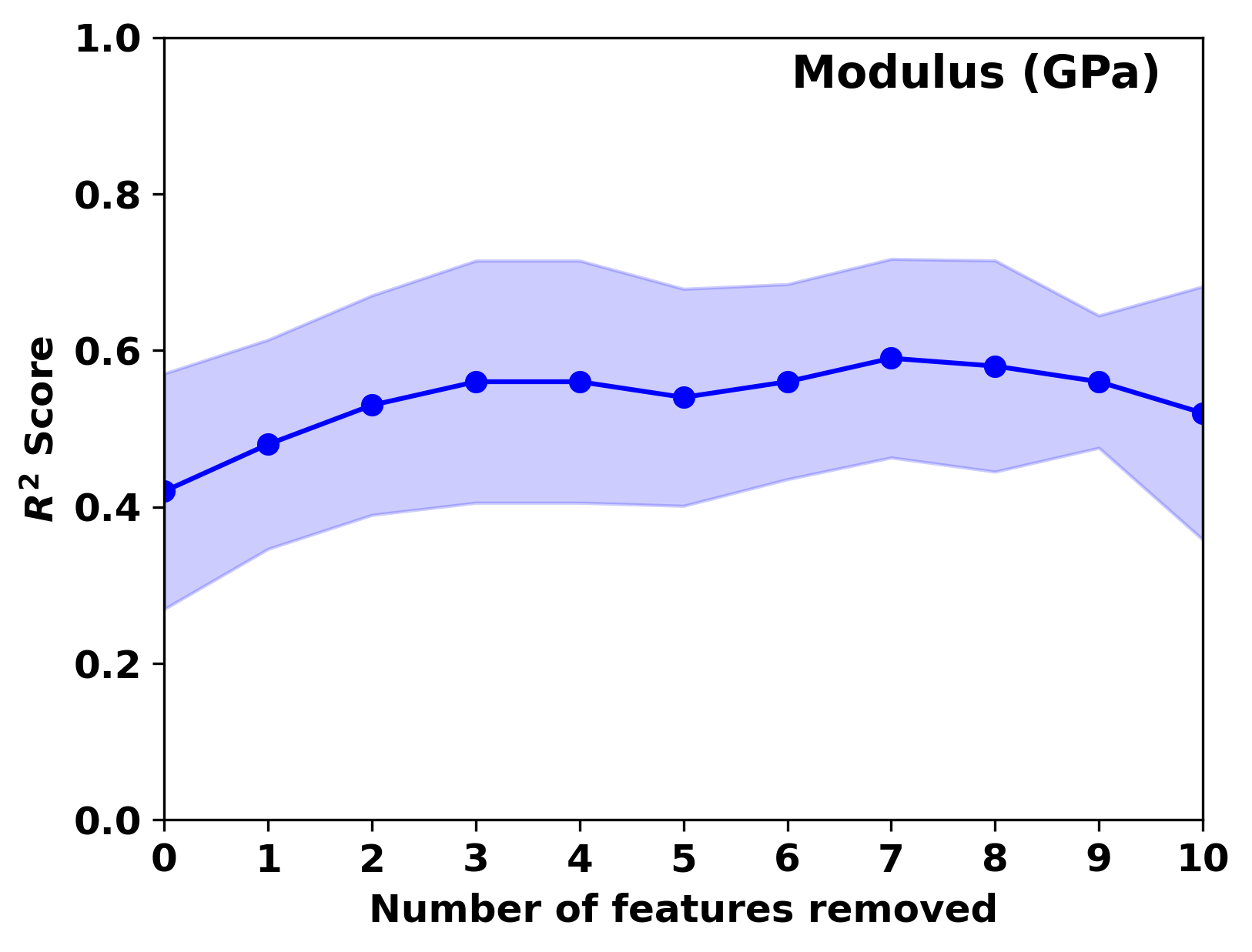}
        \caption{\label{fig:8a}}
    \end{subfigure}
    \begin{subfigure}{0.45\textwidth}
        \includegraphics [width=0.95\textwidth]{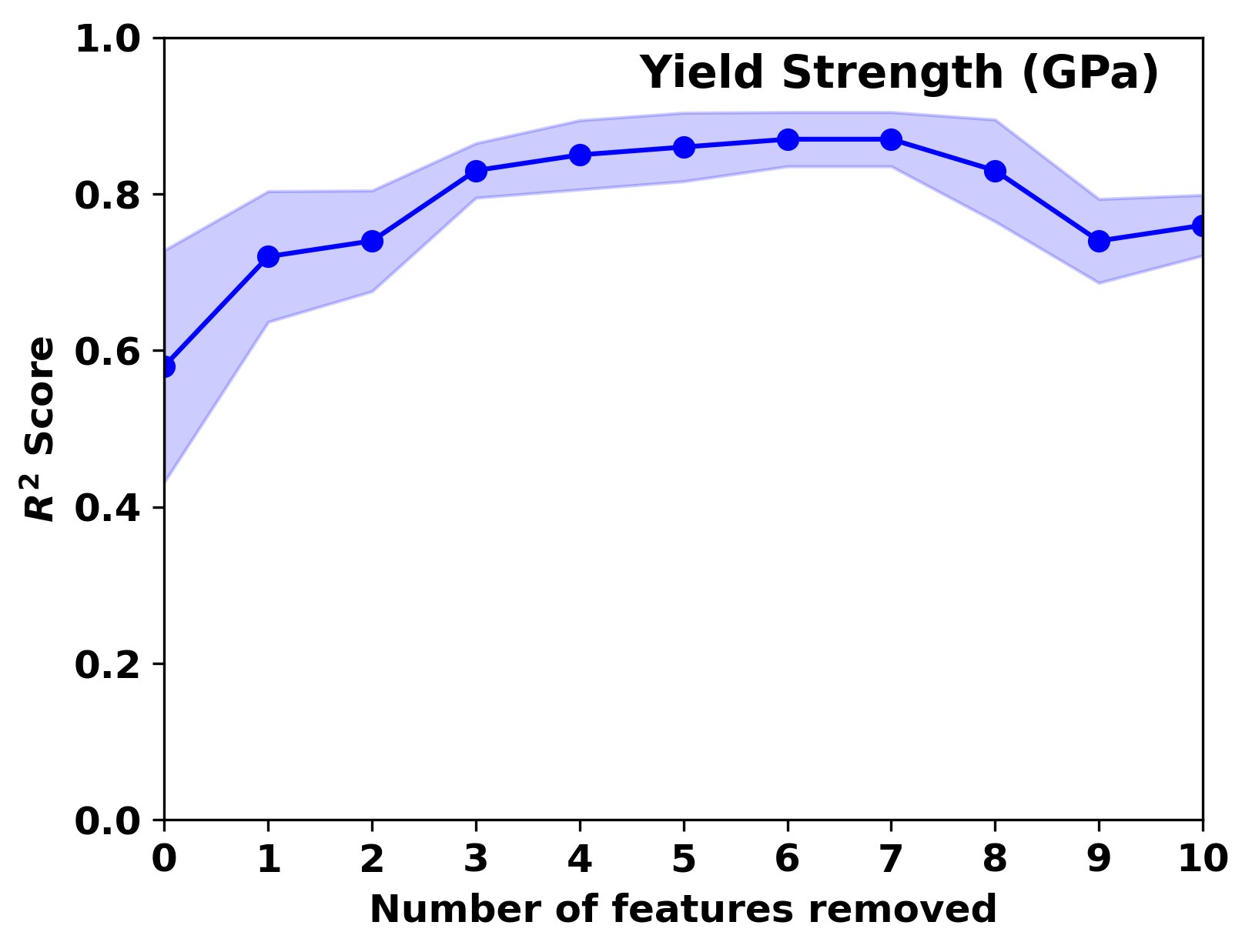}
        \caption{\label{fig:8b}}
    \end{subfigure}
    \caption{$R^2$ Score during sequential feature selection during prediction of: (\subref{fig:8a}) Elastic Modulus. (\subref{fig:8b}) Yield Strength.}
    \label{fig:8}
\end{figure}

The prediction accuracy ($R^2$ score) and details of the removed features during each trial are presented in \ref{fig:8} and Table \ref{tab:3}. We find that the optimum number of features for predicting the elastic modulus and yield strength is 6 (with maximum $R^2$ score). The important features for predicting the elastic modulus are molecular weight, fraction of carbon atoms that are SP3 hybridized, total number of carbon atoms, primary amines and secondary amines. Similarly, the important features for predicting yield strength are molecular weight, stoichiometric ratio between resin and hardener, curing percentage, total number of oxygen atoms, aromatic rings and primary amines.

\begin{table}[hbt!]
    \caption{Prediction accuracy of GPR model during backward sequential feature selection.}
    \centering
    \begin{tabular}{llll}
            \hline
            \multicolumn{2}{c}{Elastic modulus} & \multicolumn{2}{c}{Yield Strength}\\
            \hline
            Feature removed & $R^2$ Score & Feature removed & $R^2$ Score\\
            \hline
            -- &  0.42 & -- & 0.58 \\
            Oxygen atoms & 0.48 & Rotatable bonds & 0.72  \\
            Stoichiometric ratio & 0.53 & Aliphatic rings & 0.74  \\
            Valence electrons & 0.56 & Valence electrons & 0.83 \\
            Radical electrons & 0.56 & Carbon atoms & 0.85 \\
            Rotatable bonds & 0.54 & Secondary amines & 0.86  \\
            Aromatic rings & 0.56 & SP3 carbon atoms & 0.87  \\
            Aliphatic rings & 0.59 & Radical electrons & 0.87  \\
            Secondary amines & 0.58 & Stoichiometric ratio  & 0.83 \\
            Primary amines & 0.56 & Aromatic rings & 0.74  \\
            Molecular weight & 0.52 & Oxygen atoms  & 0.76 \\
            \hline
            \end{tabular}
    \label{tab:3}
\end{table}

With the selected features, we use a nested k-fold cross-validation in order to optimize the kernel function and $\alpha$ (a parameter through which the noise level of the target can be specified, the parameter also helps in dealing with numerical instabilities during fitting). In this process, we split the whole dataset into five folds, with one fold kept aside as test set and the remaining four folds taken as validation/training sets and repeat the whole process five times. We provide the following kernel functions along-with their their summations for hyperparametric tuning using grid search method: dot-product kernel [$k_D$] with initial $\sigma_0$ = 1, Mat{\'e}rn kernel [$k_M(\nu$ = 1.5)] with initial length scale, $l=1$, Gaussian kernel [$k_G$] with initial $l=1$ and constant kernel [$k_C$] with initial $C=1$ (as shown in equations \ref{eq2}-\ref{eq5}) and $\alpha$ in the range 10$^{-12}$ to 10. The terms used in Mat{\'e}rn kernel like \textit{d(.,.)}, \textit{K$_v$(.)} and $\Gamma(.)$ denote the euclidean distance, modified Bessel function and the gamma function. The parameters of kernel functions like $\sigma_0$, $l$ and $C$ are further optimized during the GPR training process.

\begin{equation}\label{eq2}
    k_D(x, y; \sigma_0)=\sigma_0^2+x \cdot y
\tag{2}
\end{equation}

\begin{equation}\label{eq3}
    k_M(x, y; \nu, l)=\frac{1}{\Gamma(\nu)2^{\nu-1}} \left (\frac{\sqrt{2\nu}}{l}d(x, y) \right)^{\nu} K_{\nu}\left (\frac{\sqrt{2\nu}}{l}d(x, y) \right)
\tag{3}
\end{equation}

\begin{equation}\label{eq4}
    k_G(x, y; l)=exp\left (-\frac{d(x, y)^2}{2l^2}\right)
\tag{4}
\end{equation}

\begin{equation}\label{eq5}
    k_C(x, y; C)= C \ \forall \ x, y
\tag{5}
\end{equation}

\begin{figure}[hbt!]
    \centering
    \begin{subfigure}{0.45\textwidth}
        \includegraphics [width=0.95\textwidth]{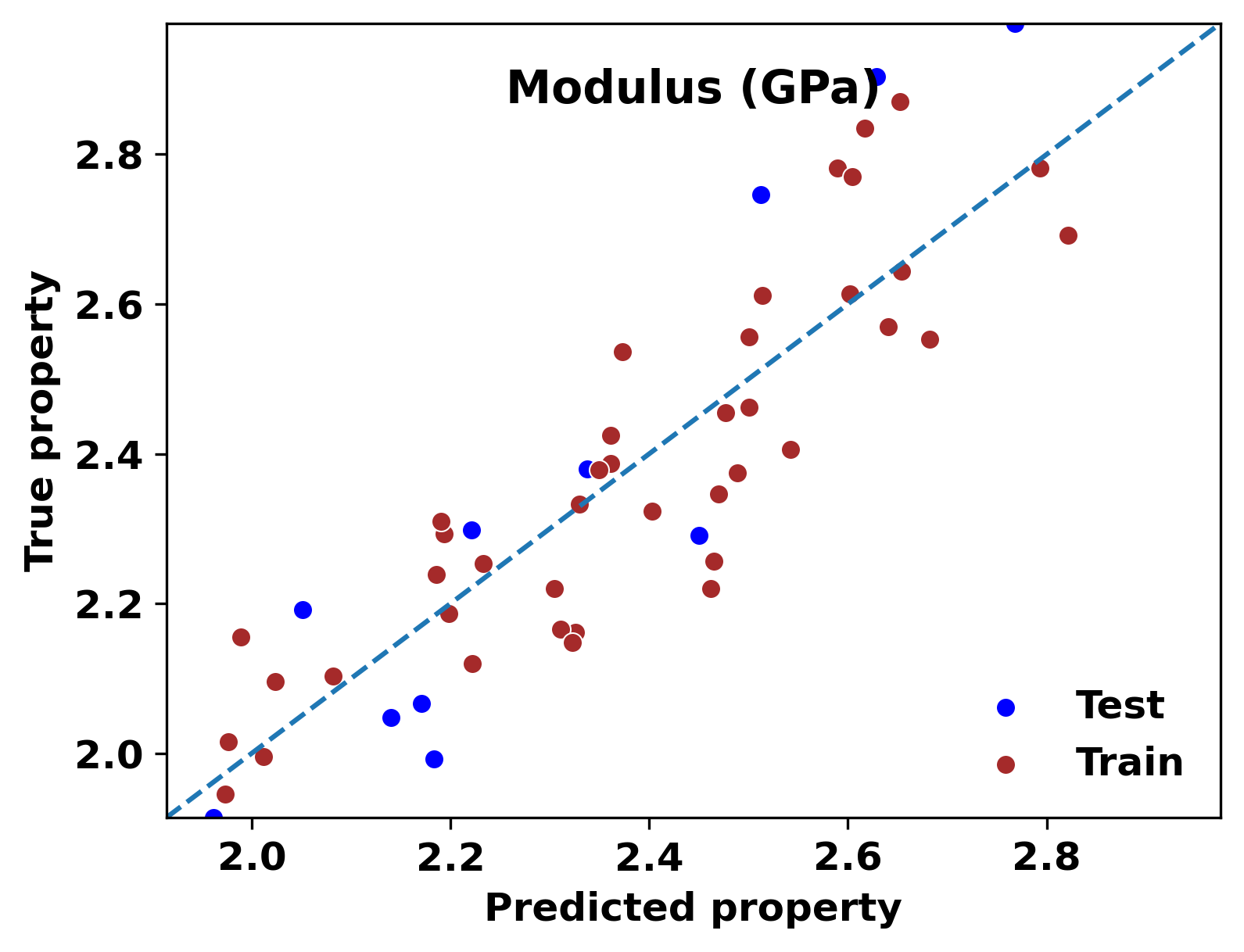}
        \caption{\label{fig:9a}}
    \end{subfigure}
    \begin{subfigure}{0.45\textwidth}
        \includegraphics [width=0.95\textwidth]{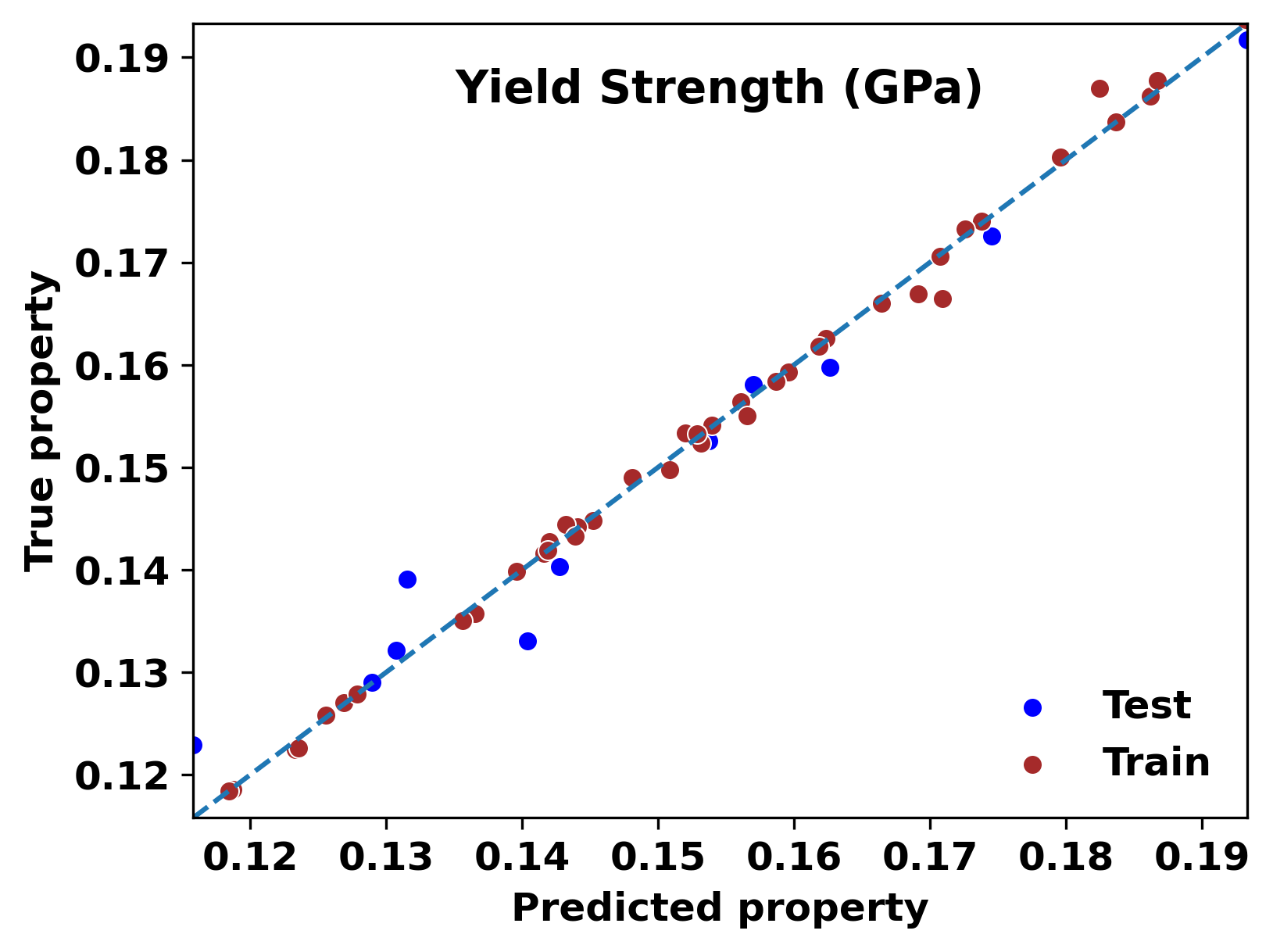}
        \caption{\label{fig:9b}}
    \end{subfigure}
    \caption{Comparison of values predicted from GPR with the reference values: (\subref{fig:9a}) Elastic Modulus (GPa). (\subref{fig:9b}) Yield strength (GPa).}
    \label{fig:9}
\end{figure}

\begin{figure}[hbt!]
    \centering
    \begin{subfigure}{0.45\textwidth}
        \includegraphics [width=0.95\textwidth]{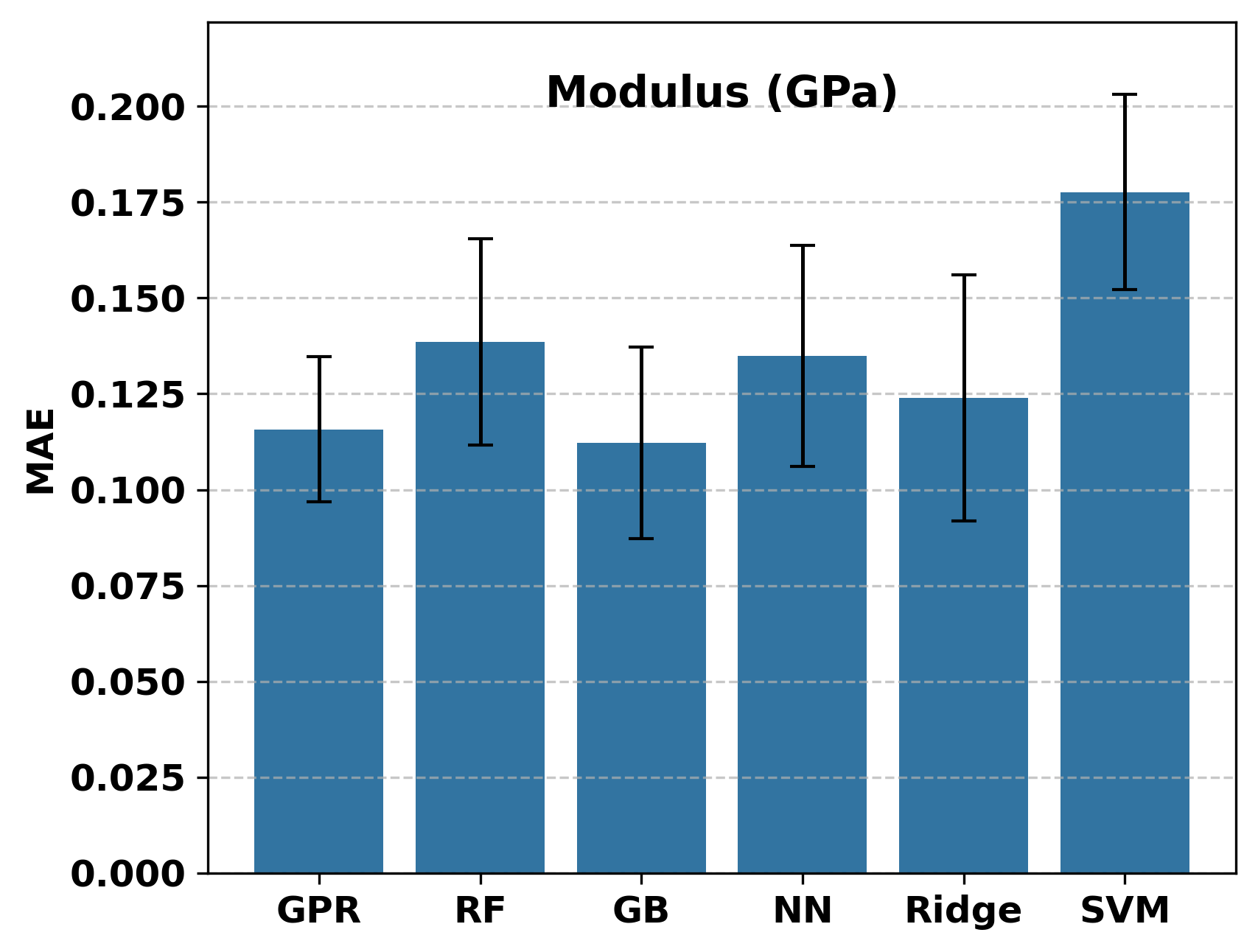}
        \caption{\label{fig:10a}}
    \end{subfigure}
    \begin{subfigure}{0.45\textwidth}
        \includegraphics [width=0.95\textwidth]{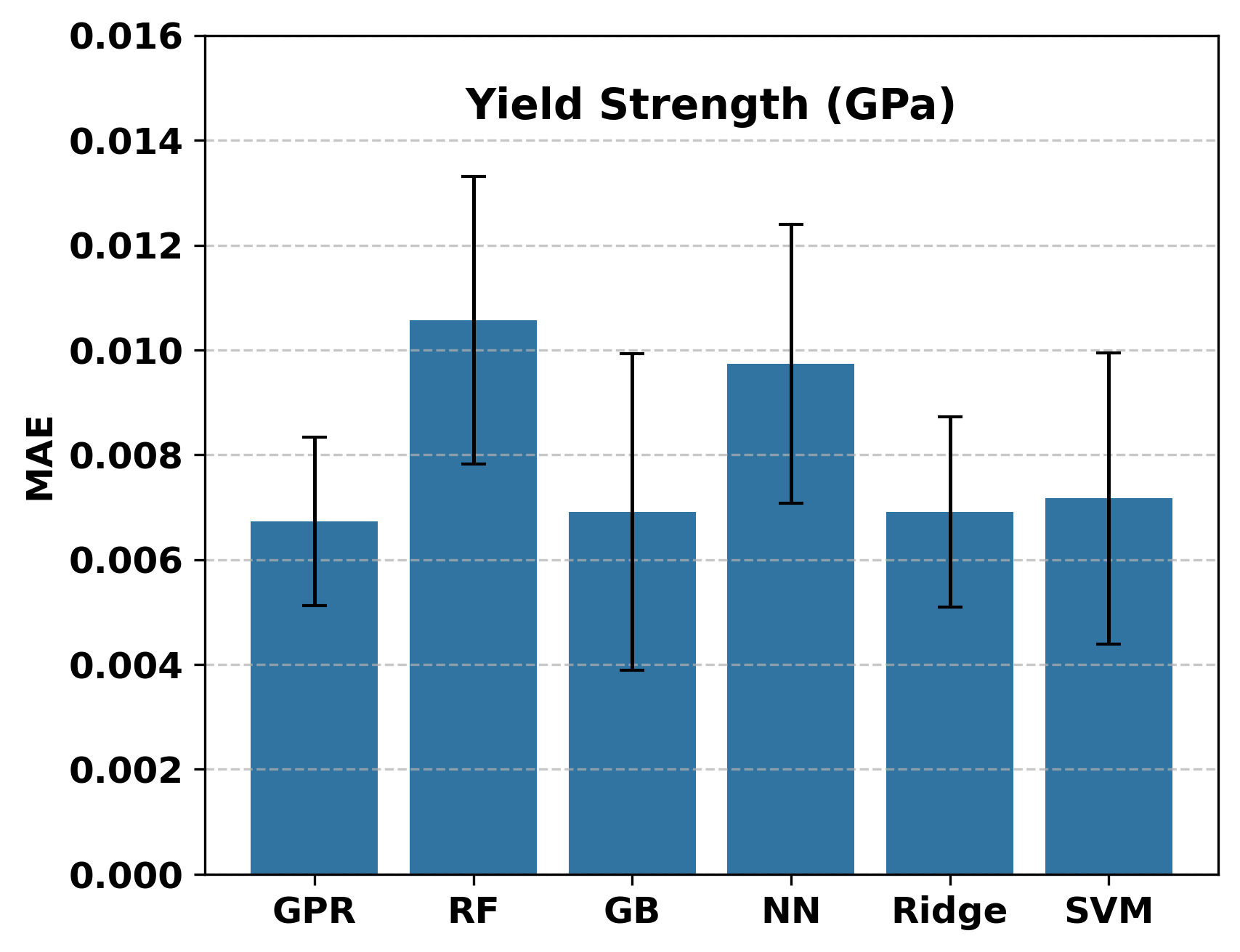}
        \caption{\label{fig:10b}}
    \end{subfigure}
    \caption{Prediction accuracy of GPR with other conventional regression models: (\subref{fig:10a}) Elastic Modulus (GPa). (\subref{fig:10b}) Yield strength (GPa).}
    \label{fig:10}
\end{figure}

We evaluate the performance of the model with different hyperparameters by measuring the $R^2$ Score between true value and predicted value on the validation set. In the case of elastic modulus, the model with summation of a constant, dot-product and Mat{\'e}rn kernel, $\alpha$ with 0.01292 is found to perform best. In the case of yield strength, the model with summation of constant and dot-product, $\alpha$ with 0.01292 is found to perform best. The accuracy of the trained GPR model with tuned kernel function and $\alpha$ for predicting elastic modulus and yield strength is presented in Table \ref{tab:4}. It can be observed that the $R^2$ Score improved from 0.42 to 0.72 in case of elastic modulus and from 0.58 to 0.81 in case of yield strength. The comparison of predicted values with the actual values both in the test set and training set is presented in Figure \ref{fig:9}.
We also compare the performance of the current model with other regression models where also feature selection using SBS and hyperparametric tuning is carried out. The MAE of the considered models is presented in Table \ref{tab:5} Figure \ref{fig:10}. It can be observed that the elastic modulus is predicted better than GPR using gradient boosting while ridge regression almost performing equivalent to GPR and the yield strength is predicted best using GPR while ridge regression, SVM and gradient boosting also seem to perform well.

\begin{table}[hbt!]
    \caption{Prediction accuracy of GPR model.}
    \centering
    \begin{tabular}{lllll}
            \hline
            Property & MAPE & MAE & RMSE & $R^2$\\
            \hline
            Elastic modulus (GPa) & 0.0481$\pm$0.0062 & 0.1157$\pm$0.0190 & 0.1393$\pm$0.0162 & 0.72$\pm$0.05\\
            Yield strength (GPa) & 0.0417$\pm$0.0086 & 0.0067$\pm$0.0016 & 0.0092$\pm$0.0024 & 0.81$\pm$0.02\\
            \hline
            \end{tabular}
    \label{tab:4}
\end{table}

\begin{table}[hbt!]
    \caption{Comparison of prediction accuracy (MAE) of GPR model with other regression models.}
    \centering
    \begin{tabular}{lllll}
            \hline
            Model & Elastic modulus (GPa) & Yield strength (GPa) \\
            \hline
            GPR & 0.1157$\pm$0.0190 & 0.0067$\pm$0.0016 \\
            Nearest Neighbor (NN) & 0.1349$\pm$0.0288 & 0.0097$\pm$0.0027 \\
            Gradient Boosting (GB) & 0.1123$\pm$0.0250 & 0.0069$\pm$0.0030 \\
            Random Forest (RF)  & 0.1349$\pm$0.0269 & 0.0106$\pm$0.0027\\
            Support Vector Machine (SVM)  & 0.1777$\pm$0.0254 & 0.0072$\pm$0.0028 \\
            Ridge Regression (Ridge) & 0.1240$\pm$0.0320 & 0.0069$\pm$ 0.0018 \\
            \hline
            \end{tabular}
    \label{tab:5}
\end{table}

\section{Conclusions} \label{sec:5}
In this study, we used molecular dynamics simulations and machine learning techniques to get in depth understanding on the structural features and process parameters affecting the mechanical properties of epoxy polymers. We used MD cross-linking simulations to develop EPON-862 based, epoxy polymeric systems with different (11) types of curing agents (hardeners) in addition to the conventionally used ones. We determined the mechanical properties (elastic modulus and yield strength) of these cross-linked polymeric systems from the MD strain simulations and validated with those from studies reported in the literature. The elastic modulus determined from MD simulations (using different strain rates) was found to be within the range of values (experimental and computational) reported in literature. The yield strength determined from MD simulations was also found to be within the range of values reported from MD simulations in the literature. However, the yield strength determined from MD simulations is found to be 3 times higher than the values reported from experimental investigations due to very high strain rate adopted in MD simulations as against the strain rate conventionally used during experiments since the strength is more sensitive to strain rate. From these simulations, we determined the influence of different parameters like the stoichiometric ratio, curing percentage and the type of hardener on the mechanical properties of epoxy polymers. We classified the hardeners into different categories like the ones with aromatic rings, aliphatic rings and aliphatic chains and the stress-strain response of each category of hardeners was analysed individually. From the analysis, it has been identified that, in the case of hardeners with aromatic rings, the hardener (DDM) with two aromatic rings possess less mechanical properties than the ones with single aromatic ring and in the case of hardeners with aliphatic chains and aliphatic rings, the hardener with more reactive amine groups possess higher mechanical properties. We further used GPR based ML model using which the mechanical properties of epoxy polymers can be predicted from their structural features extracted from cheminformatics tool. We could also determine the most important structural features whose information may be required for predicting the elastic modulus and yield strength accurately using sequential backward selection technique. We identified that parameters like molecular weight, fraction of carbon atoms that are SP3 hybridized, total number of carbon atoms, primary amines and secondary amines are most important for predicting the elastic modulus while parameters like molecular weight, stoichiometric ratio between resin and hardener, curing percentage, total number of oxygen atoms, aromatic rings and primary amines are most important for predicting the yield strength. The developed model performs with an accuracy (MAE) of 0.1157$\pm$0.0190 GPa for predicting elastic modulus and 0.0067$\pm$ 0.0016 GPa for predicting yield strength. The model also demonstrates the $R_2$ Score of 0.72$\pm$0.05 and 0.81$\pm$0.02 while predicting elastic modulus and yield strength respectively. The model has also found to perform best in comparison to conventional regression models for predicting yield strength whereas for predicting elastic modulus, it performed slightly lower than gradient boosting. The proposed method can thus be extended to simulate a larger number of epoxy polymeric systems virtually and determine their properties in an efficient way. This will further lead towards increased performance space of the epoxy polymers and also aid in designing the polymers by merely selecting the constituents with appropriate structural features.

\section*{Acknowledgments}
The authors would like to thank the help received Rick Oerder while developing the model. The first author (Sindu~B.S.) acknowledges the support in the form of Post Doctoral Industrial Fellowship received from Indo-German Science and Technology Centre (IGSTC) to carry out research in Fraunhofer Institute for Algorithms and Scientific Computing (SCAI), Germany. This paper has been assigned the registration number CSIR-SERC-1091/2023. Moreover, this work was supported in part by the BMBF-project 05M2AAA {MaGriDo} (Mathematics for Machine Learning Methods for Graph-Based Data with Integrated Domain Knowledge).
  
\bibliographystyle{IEEEtran}
\bibliography{sample}
\end{document}